\documentclass[journal,12pt,draftclsnofoot,onecolumn]{IEEEtran}
\usepackage{ifpdf}
\usepackage{mathrsfs}
\usepackage{cite}
\usepackage{amsmath}
\usepackage{amssymb}
\usepackage{stfloats}
\usepackage{graphicx}
\usepackage{setspace}
\usepackage{xcolor}

\ifCLASSOPTIONcompsoc
\usepackage[tight,normalsize,sf,SF]{subfigure}
\else
\usepackage[tight,footnotesize]{subfigure}
\fi

\begin{document}
\title{Single-Carrier Modulation for Large-Scale Antenna Systems}

\author{Yinsheng~Liu, Geoffrey Ye Li, Zhangdui Zhong, and Deli Qiao.
\thanks{Yinsheng Liu is with State Key Laboratory of Rail Traffic Control and Safety and School of Computer Science and Information Technology, Beijing Jiaotong University, Beijing 100044, China, e-mail: ys.liu@bjtu.edu.cn.}
\thanks{Geoffrey Ye Li is with ITP lab, School of ECE, Georgia Institute of Technology, Atlanta 30313, Georgia, USA, e-mail: liye@ece.gatech.edu.}
\thanks{Zhangdui Zhong is with State Key Laboratory of Rail Traffic Control and Safety, Beijing Jiaotong University, Beijing 100044, China, e-mail: zhdzhong@bjtu.edu.cn.}
\thanks{Deli Qiao is with school of IST, East China Normal University, e-mail: dlqiao@ce.ecnu.edu.cn.}
}

\maketitle
\doublespacing

\begin{abstract}

\emph{Large-scale antenna} (LSA) has gained a lot of attention due to its great potential to significantly improve system throughput. In most existing works on LSA systems, \emph{orthogonal frequency division multiplexing} (OFDM) is presumed to deal with frequency selectivity of wireless channels. Although LSA-OFDM is a natural evolution from \emph{multiple-input multiple-output} OFDM (MIMO-OFDM), the drawbacks of LSA-OFDM are inevitable, especially when used for the uplink. In this paper, we investigate \emph{single-carrier} (SC) modulation for the uplink transmission in LSA systems based on a novel waveform recovery theory, where the receiver is designed to recover the transmit waveform while the information-bearing symbols can be recovered by directly sampling the recovered waveform. The waveform recovery adopts the assumption that the antenna number is infinite and the channels at different antennas are independent. In practical environments, however, the antenna number is always finite and the channels at different antennas are also correlated when placing hundreds of antennas in a small area. Therefore, we will also analyze the impacts of such non-ideal environments.

\end{abstract}


\newpage
\section{Introduction}
\emph{Large-scale antenna} (LSA) has gained a lot of attention recently \cite{TLMarzetta,FRusek,EGLarsson,LLu}. In an LSA system, a \emph{base station} (BS) is equipped with hundreds of antennas. It can be considered as an extension of the traditional \emph{multiple-input multiple-output} (MIMO) system, which has been widely studied during the last couple of decades \cite{DTse}. The properties of LSA make it a potential technique for future wireless systems.\par

Through the employment of a large number of antennas at the BS, the channel vectors for different users will be asymptotically orthogonal if the channels corresponding to different users and different antennas are independent \cite{EGLarsson}. In this case, the \emph{matched filter} (MF) becomes the optimal detector \cite{JHoydis}. The asymptotical orthogonality of channel vectors corresponding to different users also allows multiusers to work in the same bandwidth, and thus can improve the spectrum efficiency of the network \cite{HQNgo2}. It is shown that the transmit power for each user can be scaled down by the antenna number or the square root of the antenna number, depending on whether accurate channel parameters are available or not \cite{HQNgo}. As a result, the transmit powers of users can be arbitrarily small when the number of antennas is large enough, and the energy efficiency of users can be therefore significantly improved.\par

Traditional MIMO systems are usually combined with \emph{orthogonal frequency division multiplexing} (OFDM), where the latter is used to convert frequency selective channels into flat fading channels. As a natural evolution of MIMO-OFDM, OFDM has been presumed in most existing works in LSA systems \cite{EGLarsson,JHoydis,HQNgo}. Although straightforward, the drawbacks of LSA-OFDM are inevitable, especially when used for the uplink. The high \emph{peak-to-average power ratio} (PAPR) of OFDM will reduce the efficiency of the power amplifier at the user terminal, making OFDM not suitable for uplink transmission. Meanwhile, the BS will need hundreds of discrete Fourier transforms for OFDM demodulation, resulting in a heavy computational burden. On the other hand, it is well known that \emph{single carrier} (SC), due to its low PAPR, has replaced OFDM for the uplink of \emph{long-term evolution} (LTE) even though the latter has been widely advocated for the downlink. Following the same philosophy as LTE uplink, SC should be still favorable for the uplink in LSA systems to save the needs of high performance power amplifiers at the user terminals. In view of the advantages of SC over OFDM when used for the uplink, LSA-SC is expected to be a competitive candidate for uplink transmission in future wireless systems.\par

Due to the reasons above, SC has been investigated in LSA systems, especially for the uplink. In \cite{YLiu}, we have considered SC for the uplink transmission in an LSA system over Rician fading channels. In that case, an equalizer is required at the receiver to suppress multiuser interference caused by the line-of-sight path. In \cite{APitar}, SC is used for precoding with an MF precoding matrix. The work in \cite{APitar} can be easily extended to uplink according to the duality \cite{DTse}. Traditionally, the optimal multi-antenna SC receiver is composed of a set of analog MFs followed by an equalizer that is used to suppress the \emph{inter-symbol interference} (ISI) \cite{GLi,PBalaban,PBalaban2}. The need of the equalizer significantly increases the complexity of the SC receiver due to the matrix inverse. This is especially the case when SC is used for high-speed data transmission.\par

In this paper, we propose an LSA-SC receiver where the ISI is suppressed through the LSA array. The LSA-SC receiver is only composed of a set of analog MFs and the equalizer is not required any more. Hence, the complexity of LSA-SC receiver can be greatly reduced compared to the traditional multi-antenna SC receiver \cite{GLi,PBalaban}. Essentially, the LSA-SC receiver in this paper can be viewed as a special case of the traditional multi-antenna SC receiver when the antenna number is infinite and the channels at different antennas are independent \cite{GLi,PBalaban}. As an alternative, our derivation is based on a novel waveform recovery theory, where the receiver is designed to to recover the transmit waveform while the information-bearing symbols can be recovered by directly sampling the recovered waveform. Compared to existing works \cite{GLi,PBalaban}, it provides a much simpler way to understand the behavior of SC receiver in large-scale regime. On the other hand, the waveform recovery theory adopts the assumption that the antenna number is infinite and the channels at different antennas are independent. In practical environments, however, the antenna number is always finite and the channels at different antennas will be correlated when placing hundreds of antennas in a small area. We will therefore analyze the impact of the non-ideal environments in this paper.\par

The rest of this paper is organized as follows. Section II introduces the system model. In Section III, we investigate the waveform recovery in an LSA-SC system. In Section IV, we will discuss the impact of non-ideal environments on the LSA-SC receiver. Numerical results are presented in Section V. Finally, conclusion is drawn in Section VI.

\section{System Model}
In this section, we will first introduce the channel model, and then we will present SC modulation for an LSA system.
\subsection{Channel Model}

For a frequency-selective fading channel, the basedband \emph{channel impulse response} (CIR) at the $m$-th receive antenna can be given by
\begin{align}\label{2-1}
c_m(t)=\sum_l\alpha_m[l]\delta(t-\tau_l),
\end{align}
where $\alpha_m[l]$ is the complex gain of the $l$-th tap on the $m$-th antenna, $\tau_l$ is the corresponding tap delay, and $\delta(\cdot)$ represents the Dirac delta function. In (\ref{2-1}), we have implicitly assumed that the tap delays for different antennas are the same since the extra propagation delays caused by the physical size of the antenna array are quite small and thus can be omitted. If the gains corresponding to different taps are independent and complex Gaussian, then
\begin{align}\label{2-2}
\mathrm{E}\left\{\alpha_m[l]\alpha_m^*[p]\right\}=\sigma_l^2\delta[l-p],
\end{align}
where $\sigma_l^2$ is the power of the $l$-th tap and $\delta[\cdot]$ denotes the Kronecker delta function.\par

Denote $\mathbf{c}(t)=\left[c_1(t),\cdots,c_M(t)\right]^{\mathrm{T}}$ to be the CIR vector corresponding to different antennas at the BS. From (\ref{2-1}), we have
\begin{align}\label{2-3}
\mathbf{c}(t)=\sum_{l}\boldsymbol{\alpha}[l]\delta(t-\tau_l),
\end{align}
where $\boldsymbol{\alpha}[l]=\left(\alpha_1[l],\cdots,\alpha_M[l]\right)^{\mathrm{T}}$ denotes the complex gain vector corresponding to the $l$-th tap. If assuming the antenna number in an LSA system is infinite and the complex gains at different antennas are independent, the mean in (\ref{2-2}) is then equal to the sample mean according to the law of large numbers \cite{APapoulis}, that is
\begin{align}\label{2-3-0}
\lim\limits_{M\to \infty}\frac{1}{M}\boldsymbol{\alpha}^{\mathrm{H}}[l]\boldsymbol{\alpha}[p]&=\mathrm{E}\left\{\alpha_m[l]\alpha_m^*[p]\right\}\nonumber\\
&=\sigma_l^2\delta[l-p],
\end{align}
since $\alpha_l[m]$'s with $m=1,2,\cdots,M$ are independent and identically distributed random variables. The last identity indicates that the complex gain vectors corresponding to different taps will be asymptotically orthogonal.

\subsection{SC Modulation}

\begin{figure}
  \centering
  \includegraphics[width=5in]{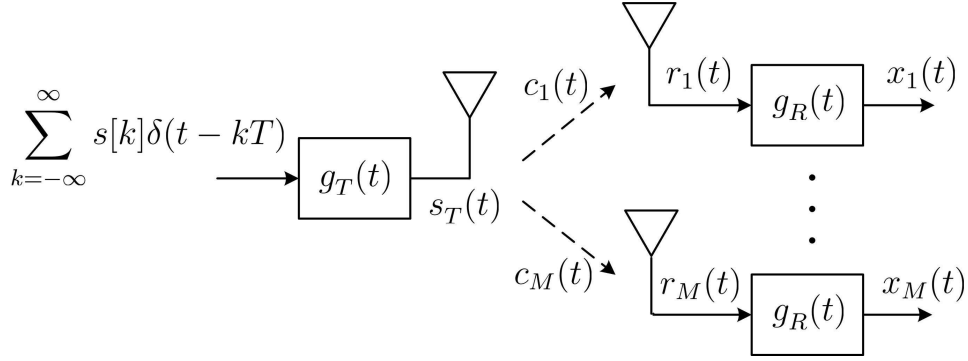}\\
  \caption{SC transmission with multiple receive antennas.}\label{sc_model}
\end{figure}

Fig.~\ref{sc_model} shows the uplink SC transmission in an LSA system with $M$ receive antennas. Without loss of generality, we assume that the user is equipped with only one transmit antenna for simplicity even though our result can be directly applied to the multiple antenna case. If the transmitted symbols, $s[k]$'s, are independent with zero mean and unit variance, then we have $\mathrm{E}\left\{s[k]s^*[n]\right\}=\delta[k-n]$. Given a transmit filter, $g_T^{}(t)$, the transmit waveform is
\begin{align}
s_T^{}(t)=\sum_{k=-\infty}^{\infty}s[k]g_T^{}(t-kT),
\end{align}
where $T$ is the symbol period.\par
Denote $\mathbf{r}(t)=\left[r_1(t),\cdots,r_M(t)\right]^{\mathrm{T}}$ to be the received signal vector at the BS, then we have
\begin{align}
\mathbf{r}(t)=\sum_l\boldsymbol{\alpha}[l]s_T^{}(t-\tau_l) + \mathbf{v}(t),
\end{align}
where $\mathbf{v}(t)=[v_1(t),\cdots,v_M(t)]^{\mathrm{T}}$ is the \emph{additive white Gaussian noise} (AWGN) with a constant spectral density, $N_0$. After going through a receive filter, $g_R(t)$, at each antenna, the filtered receive signal vector, $\mathbf{x}(t)=\left[x_1(t),\cdots,x_M(t)\right]^{\mathrm{T}}$, becomes
\begin{align}\label{2-3-1}
\mathbf{x}(t)&=\mathbf{r}(t) * g_R^{}(t)\nonumber\\
&=\sum_l\boldsymbol{\alpha}[l]s(t-\tau_l) + \mathbf{z}(t),
\end{align}
where $s(t)$ is given by
\begin{align}
s(t)=\sum_{k=-\infty}^{\infty}s[k]g(t-kT),
\end{align}
with $g(t)=g_T^{}(t) * g_R^{}(t)$ and $\mathbf{z}(t)=\left[z_1(t),\cdots,z_M(t)\right]^{\mathrm{T}}$ is the filtered noise vector with $z_m(t)=v_m(t) * g_R^{}(t)$.\par
In general, the same root-raised-cosine function with roll-off factor, $\beta$, is used for both transmit and receive filters, and the overall impulse response thus becomes a raised-cosine function which can satisfy Nyquist condition \cite{JGProakis}, that is, $g(kT)=\delta[k]$ and thus $s(kT)=s[k]$. In this case, the correlation matrix of the filtered noise can be expressed by
\begin{align}
\mathrm{E}\{\mathbf{z}(t)\mathbf{z}^{\mathrm{H}}(t+\tau)\}=N_0g(\tau)\mathbf{I},
\end{align}
where $\mathbf{I}$ denotes the identity matrix.

\section{LSA-SC System}
In this section, we will first introduce the single-tap LSA-SC receiver using waveform recovery, then a multi-tap receiver will be addressed. Finally, the relationship with the traditional multi-antenna SC receiver will be discussed.
\subsection{Single-Tap Receiver}
\begin{figure}
  \centering
  \includegraphics[width=2.6in]{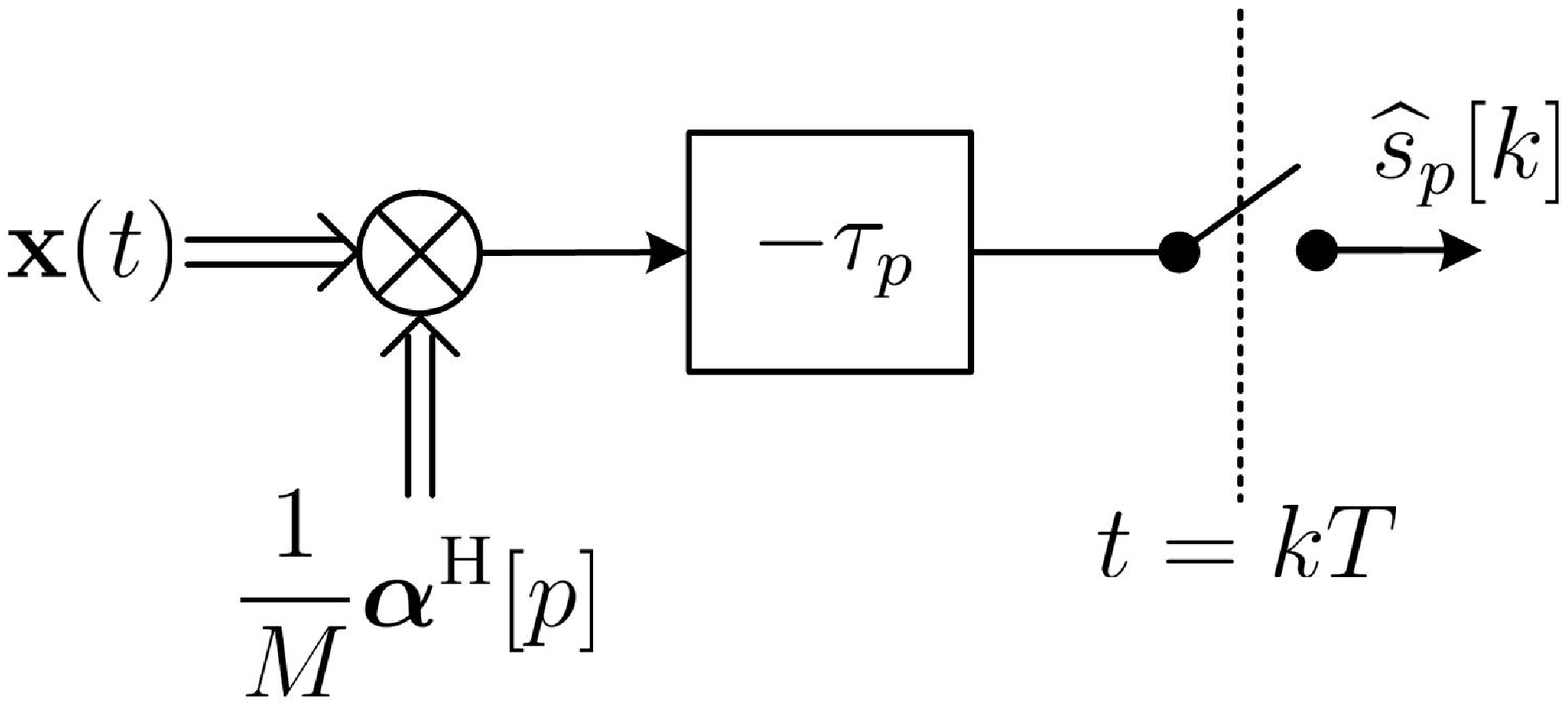}~~~~
  \includegraphics[width=3.2in]{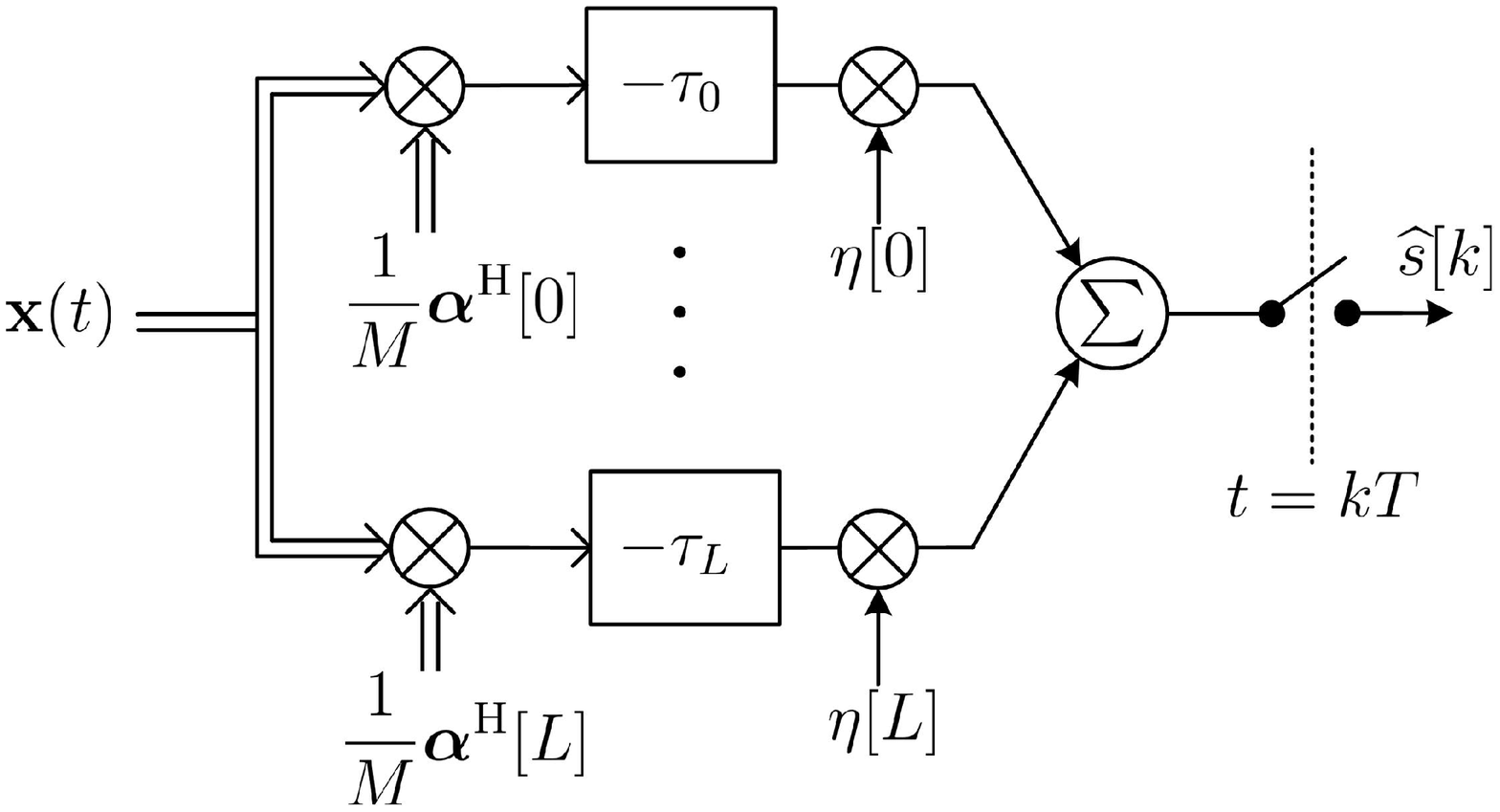}\\
  (a)~~~~~~~~~~~~~~~~~~~~~~~~~~~~~~~~~~~~~~~~~~~~(b)
  \caption{LSA-SC receiver with (a) single-tap (b) multi-tap.}\label{paths}
\end{figure}

As shown in Fig.~\ref{paths} (a), a single-tap receiver exploits a particular tap from the received signal to recover the desired waveform, $s(t)$. To recover the waveform over the $p$-th tap with delay $\tau_p$, as in Fig.~\ref{paths} (a), we have
\begin{align}\label{3-1}
\widehat{s}_p(t)=\frac{1}{M}\boldsymbol{\alpha}^{\mathrm{H}}[p]\mathbf{x}(t+\tau_p),
\end{align}
Using the asymptotical orthogonality in (\ref{2-3-0}), the complex gain vectors are asymptotically orthogonal if assuming the antenna number is infinite and the channels at different antennas are independent. In this case, (\ref{3-1}) can be obtained as
\begin{align}\label{3-2-0}
\widehat{s}_p(t)=\sigma_p^2 s(t) + w_p(t),
\end{align}
which is composed of the desired waveform, $s(t)$, except for being scaled by a factor, $\sigma_p^2$, and the additive noise, $w_p(t)$, with
\begin{align}
w_p(t)=\frac{1}{M}\boldsymbol{\alpha}^{\mathrm{H}}[p]\mathbf{z}(t+\tau_p),
\end{align}
whose correlation function is given by
\begin{align}
\mathrm{E}\{w_p(t)w_p^*(t+\tau)\}=\frac{N_0}{M}g(\tau)\sigma_p^2.
\end{align}
From (\ref{3-2-0}), the multipath propagation has been completely removed using the asymptotical orthogonality in (\ref{2-3-0}). The desired waveform can be therefore recovered over the tap of interest except for a scaling factor.\par

Note that the waveform recovery theory only provides a way to understand the SC receiver in large-scale regime. The purpose of a communication system is still to recover the information-bearing symbols. In this sense, a sampler is required for the recovered waveform, as shown in Fig.~\ref{paths} (a). Since $g(t)$ satisfies the Nyquist condition, the symbols can be easily recovered with baud-rate sampling. Therefore, the decision variable corresponding to the $p$-th tap at $t=kT$ is
\begin{align}
\widehat{s}_p[k]=\sigma_p^2 s[k] + w_p(kT).
\end{align}
Accordingly, the \emph{signal-to-noise ratio} (SNR) corresponding to the $p$-th  tap will be
\begin{align}\label{3-3}
\mathrm{SNR}_p=\sigma_p^2\frac{M}{N_0}.
\end{align}

\subsection{Multi-Tap Receiver}
In above, the desired waveform is recovered with only the $p$-th tap. Actually, the SNR can be further improved by exploiting all taps through a linear combination of $\widehat{s}_p(t)$'s with $p=0,1,\cdots,L$, as shown in Fig.~\ref{paths} (b).\par
Let $\eta[p]$ with $p=0,1,\cdots,L$ be the coefficients for the linear combination, then the overall recovered waveform from all taps can be expressed as
\begin{align}\label{3-3-0}
\widehat{s}(t)&=\sum_{p=0}^{L}\eta[p]\widehat{s}_p(t)\nonumber\\
&=\sum_{p=0}^{L}\eta[p]\sigma_p^2s(t)+\sum_{p=0}^{L}\eta[p]w_p(t).
\end{align}
Similar to the single-tap case, the decision variable after sampling at $t=kT$ can be obtained as
\begin{align}\label{3-3-1}
\widehat{s}[k]=\sum_{p=0}^{L}\eta[p]\sigma_p^2s[k]+\sum_{p=0}^{L}\eta[p]w_p(kT).
\end{align}
Accordingly, the SNR after combination will be
\begin{align}\label{3-4}
\mathrm{SNR}=\frac{M\displaystyle{\left(\sum_{p=0}^L\sigma_p^2\eta[p]\right)^2}}{N_0\displaystyle{\sum_{p=0}^L\sigma_p^2\eta^2[p]}}.
\end{align}
Direct calculation from (\ref{3-4}) yields that the maximum SNR can be achieved when
\begin{align}\label{3-5}
\eta[0]=\eta[1]=\cdots=\eta[L],
\end{align}
and the maximum SNR, if we assume that the total channel power is unit, will be
\begin{align}\label{3-6}
\mathrm{SNR}=\frac{M}{N_0}\geq \mathrm{SNR}_p,
\end{align}
for any $p$. As expected, the SNR of the multi-tap receiver can be improved compared to the single-tap receiver.

\subsection{Relation to Traditional Multi-Antenna Receiver}

\begin{figure}
  \centering
  \includegraphics[width=4.5in]{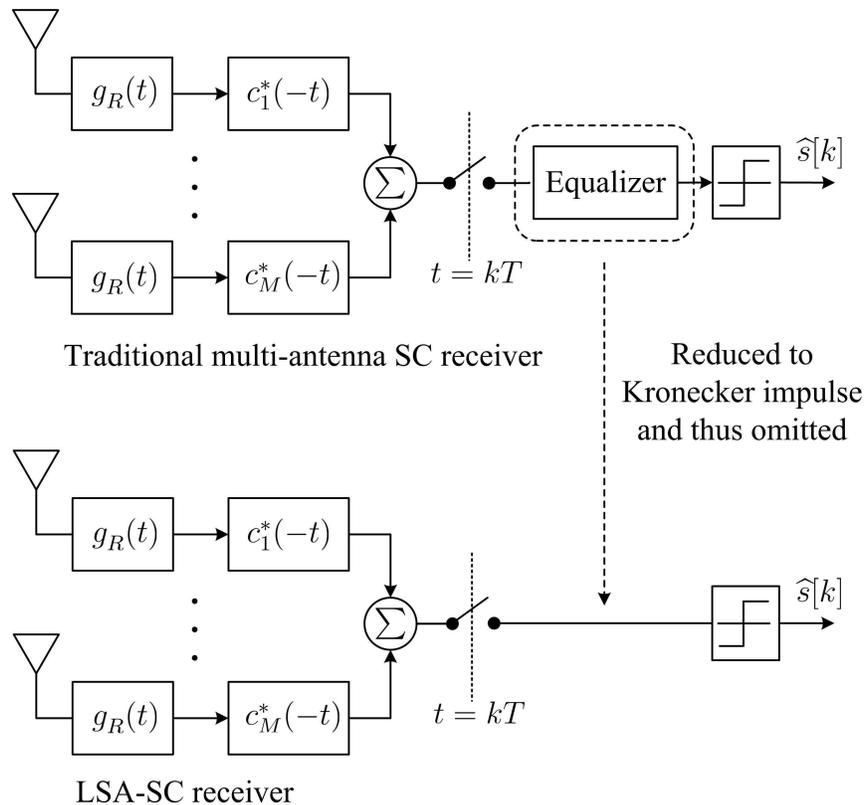}\\
  \caption{From traditional multi-antenna SC receiver to LSA-SC receiver. The Equalizer reduces to a Kronecker impulse and thus can be omitted in the LSA-SC receiver.}\label{LSA_SC}
\end{figure}

Without loss of generality, we assume that $\eta[p]=1$ for simplicity. Then, by substituting (\ref{3-1}) into (\ref{3-3-0}), we can obtain
\begin{align}\label{2-12}
\widehat{s}(t)=\frac{1}{M}\boldsymbol{c}^{\mathrm{H}}(-t)*\mathbf{x}(t),
\end{align}
where (\ref{2-1}) has been used for this equation. The receiver structure for (\ref{2-12}) is shown as in Fig.~\ref{LSA_SC}. From the figure, the LSA-SC receiver is only composed of a set of analog filters and the equalizer is not required any more. Hence, the complexity of LSA-SC receiver can be substantially reduced compared to the traditional multi-antenna SC receiver \cite{PBalaban}.\par

Essentially, the LSA-SC receiver can be viewed as a special case of the traditional multi-antenna SC receiver if the antenna number is infinite and the channels at different antennas are independent. In that case, the equalizer in traditional multi-antenna SC receiver will reduce to a Kronecker impulse, and thus can be omitted in the receiver. Although the resulted receivers are similar, the waveform recovery theory based on the asymptotical orthogonality in LSA systems provides a straightforward way to understand the behavior of SC in large-scale regime. This is especially the case when the LSA-SC receiver works in non-ideal environment as it will be discussed in Section IV. \par

\section{Impact of Non-Ideal Environments}
The derivation of the LSA-SC receiver in Section III assumes that the antenna number is infinite and the channels at different antennas are independent. In practical systems, however, the antenna number is always finite and the channels at different antennas will be also correlated when placing hundreds of antennas in a small area. Therefore, we will investigate the impact of non-ideal environments on the LSA-SC system in this section.\par
\begin{figure}
  \centering
  \includegraphics[width=3in]{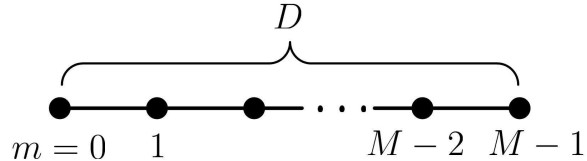}\\
  \caption{ULA: antennas are placed along a line with length $D$.}\label{ULA}
\end{figure}
\subsection{Spatial Correlation}
Without loss of generality, we consider a typical \emph{uniform-linear-array} (ULA), where $M$ antennas are placed along a line with length $D$ as in Fig.~\ref{ULA}. The spatial correlation analysis here is also valid for any other form of antenna arrays, and the correlation function will then depend on the geometry of the array. In the ULA, the distance between the $m$-th and the $m_0$-th antennas will be $d=(m-m_0)D/(M-1)$. Denote the spatial correlation between any two antennas with distance, $d$, to be
\begin{align}
\rho\left(d/\lambda\right)=\rho\left(\frac{m-m_0}{M-1}\cdot\frac{D}{\lambda}\right).
\end{align}
The particular form of $\rho(\cdot)$ depends on the surrounded environments. If the power azimuth spectrum is a constant from $0$ to $2\pi$ \cite{LSchumacher}, then
\begin{align}\label{5-1}
\rho\left(d/\lambda\right)=J_0\left(2\pi d/\lambda\right),
\end{align}
where $J_0(\cdot)$ is the zeroth-order Bessel function of the first kind.\par
In the presence of spatial correlation, we have
\begin{align}\label{5-2}
\mathrm{E}\left\{\alpha_m[l]\alpha_{m_0}^*[p]\right\}=\sigma_l^2\rho\left(\frac{m-m_0}{M-1}\cdot\frac{D}{\lambda}\right)\delta[l-p].
\end{align}
Accordingly, the correlation matrix for the tap vectors, $\boldsymbol{\alpha}[l]$, is given by
\begin{align}\label{5-2-0}
\mathrm{E}\left\{\boldsymbol{\alpha}[l]\boldsymbol{\alpha}^{\mathrm{H}}[p]\right\}=\sigma_l^2\delta[l-p]\mathbf{R},
\end{align}
where $\mathbf{R}$ is the spatial correlation matrix, with $(m,m_0)$-th entry being
\begin{align}
\left[\mathbf{R}\right]_{(m,m_0)}=\rho\left(\frac{m-m_0}{M-1}\cdot\frac{D}{\lambda}\right).
\end{align}
Since the channels at different antennas are correlated, the asymptotical orthogonality in (\ref{2-3-0}) is not valid any more.

\subsection{Residual ISI}
If the antenna number is finite or the channels at different antennas are correlated, the recovered waveform can be obtained, from (\ref{2-12}), as
\begin{align}\label{5-3}
\widehat{s}(t)=\frac{1}{M}\sum_p\|\boldsymbol{\alpha}[p]\|_2^2 s(t)+\frac{1}{M}\sum_p\sum_{p\neq l}\boldsymbol{\alpha}^{\mathrm{H}}[p]\boldsymbol{\alpha}[l]s(t-\tau_l+\tau_p)+\sum_p w_p(t),
\end{align}
where the first term in (\ref{5-3}) is the desired waveform, and the second term is the \emph{inter-wave interference} (IWI) which denotes the interference to the desired waveform caused by the multipath propagation. If the antenna number is infinite and the complex tap gains corresponding to different antennas are independent, the IWI can be completely suppressed using the asymptotical orthogonality, as we have discussed in Section III. However, it cannot be completely removed in the case of finite antenna number or in the presence of spatial correlation, resulting in residual ISI to the decision variable after sampleing. Note that
the IWI is in general different from the ISI. They are only equivalent when tap delays are integer times of symbol period.\par

To analyze the residual ISI, we rewrite the received symbol, from Fig.~\ref{LSA_SC}, in a symbol spaced form as
\begin{align}\label{5-3-0}
\widehat{s}[k]={f[0]s[k]}+{\sum_{n\neq 0}f[n]s[k-n]}+w[k],
\end{align}
where $f[n]$ is the overall discrete impulse response observed by the receiver, that is,
\begin{align}
f[n]=\frac{1}{M}\sum_l\sum_p\boldsymbol{\alpha}^{\mathrm{H}}[p]\boldsymbol{\alpha}[l]g(nT-\tau_l+\tau_p),\label{5-4}
\end{align}
and $w[k]$ is given by
\begin{align}
w[k]=\sum_{p}w_p(kT).
\end{align}
In (\ref{5-3-0}), the first term is the desired symbol, the second term is the residual ISI, and the third term is the additive noise.\par

In the ideal case, that is, the antenna number $M\rightarrow \infty$ and the channels at different antennas are independent, the complex gain vectors for different taps will be asymptotically orthogonal and thus $f[n]=\delta[n]$. As a result, the ISI in (\ref{5-3-0}) can be completely suppressed.\par

In the non-ideal environments, the asymptotical orthogonality in (\ref{2-3-0}) is unavailable. In this case, $f[n]\neq \delta[n]$ and thus ISI arises. The residual ISI can be expressed as
\begin{align}
I[k]=\sum_{n\neq k}f[n]s[k-n].
\end{align}
Direct calculation from the Appendix shows that the average power of residual ISI can be expressed as
\begin{align}\label{6-2}
P_{\mathrm{ISI}}&=\mathrm{E}\left\{|I[k]|^2\right\}\nonumber\\
&=P_0\sum_{n\neq 0}\sum_{p}\sum_l\sigma_p^2\sigma_l^2g^2(nT-\tau_l+\tau_p),
\end{align}
where
\begin{align}\label{6-2-0}
P_0=\frac{\mathrm{tr}\{\mathbf{R}^2\}}{M^2}.
\end{align}
In (\ref{6-2}), $P_{\mathrm{ISI}}$ is determined by the spatial correlation through $P_0$ and the power delay profile. To get more insights, we consider the following three different cases.
\subsubsection{Finite Antenna Number}
In this case, we have $\mathbf{R}=\mathbf{I}$ since the channels at different antennas are independent, but $P_0=1/M$ rather than zero since the antenna number is finite. As a result, (\ref{6-2}) reduces to
\begin{align}\label{6-2-1}
P_{\mathrm{ISI}}=\frac{1}{M}\sum_{n\neq 0}\sum_{p}\sum_l\sigma_p^2\sigma_l^2g^2(nT-\tau_l+\tau_p),
\end{align}
From (\ref{6-2-1}), the power of residual ISI is inverse proportional to the number of antennas. In an extreme case, we have $P_{\mathrm{ISI}}=0$ if $M\rightarrow \infty$, which agrees with our analysis in ideal environments.\par
\subsubsection{Spatial Correaltion}
If the channels at different antennas are correlated, $\mathrm{tr}\{\mathbf{R}^2\}$ can be rewritten, from the Appendix, as
\begin{align}
\mathrm{tr}\{\mathbf{R}^2\} = \sum_{m=-(M-1)}^{M-1}\rho^2\left(\frac{m}{M-1}\cdot\frac{D}{\lambda}\right)(M-|m|).
\end{align}
Therefore, when $M\rightarrow\infty$, $P_0$ can be rewritten as
\begin{align}\label{6-3}
P_0 = \int_{-1}^{+1}\rho^2\left(\frac{x D}{\lambda}\right)(1-|x|)dx>0.
\end{align}
It means that in the presence of spatial correlation, the ISI cannot be completely suppressed even when the antenna number is large enough.\par
In addition, direct calculation of (\ref{6-3}) for the typical Jakes correlation in (\ref{5-1}) yields that \cite{IGradshteyn}
\begin{align}
P_0&=2\cdot{}_2F_3\left(\frac{1}{2},\frac{1}{2};1,1,\frac{3}{2};-\frac{4\pi^2 D^2}{\lambda^2}\right)-J_0^2\left(\frac{2\pi D}{\lambda}\right)-J_1^2\left(\frac{2\pi D}{\lambda}\right)\nonumber\\
&\approx 2\cdot{}_2F_3\left(\frac{1}{2},\frac{1}{2};1,1,\frac{3}{2};-\frac{4\pi^2 D^2}{\lambda^2}\right),
\end{align}
where ${}_2F_3(\cdot;\cdot;\cdot)$ is the generalized hypergeometric function and $J_1(\cdot)$ denotes the first order Bessel function of the first kind. The last identity is due to the fact that $J_0(\cdot)$ and $J_1(\cdot)$ are very small when $D/\lambda$ is large enough. Since ${}_2F_3(\cdot;\cdot;\cdot)$ increases as the decreasing of $D/\lambda$, the ISI power can be therefore reduced by using longer ULA.
\subsubsection{Integer Tap Delays}
In this case, the tap delays are integer times of symbol duration, that is, $\tau_l=lT$. Then, we have
$g^2(nT-\tau_l+\tau_p)=\delta[n - l+p]$, and substituting this identity into (\ref{6-2}), we can obtain
\begin{align}\label{6-5-0}
P_{\mathrm{ISI}}=P_0\left(1-\sum_{l=0}^{L}\sigma_l^4\right).
\end{align}
Apparently, $P_{\mathrm{ISI}}$ in (\ref{6-5-0}) can achieve its maximum when $\sigma_l^2$'s are all equal if assuming $\sum_l\sigma_l^2=1$. In this case, we have
\begin{align}
P_{\mathrm{ISI,max}}=\frac{L}{L+1}P_0\approx P_0,
\end{align}
where the second approximation holds on when $L$ is large. It means that the maximum of the residual ISI power is irrelevant to the power delay profile but entirely determined by the spatial correlation

\section{Simulation Results}
In this section, numerical results are presented to evaluate the proposed LSA-SC system. In the simulation, we considered a \emph{quadrature-phase-shift-keying} (QPSK) modulated SC signal with roll-off factor $\beta=0.25$ and symbol duration $T=0.2~\mu\text{s}$. The BS is equipped a ULA with $M=100$ antennas without specification. A normalized \emph{extended typical urban} (ETU) channel model \cite{3GPP_36104} is used in the simulation, which has $9$ taps with maximum delay spread $5~\mu\text{s}$. Both spatially independent and correlated channels are considered in the simulation. For practical implementation of the LSA-SC receiver, we adopt a twice oversampling based approach where the analog MF can be implemented in a discrete manner \cite{GLi}.

\begin{figure}
  \centering
  \includegraphics[width=4in]{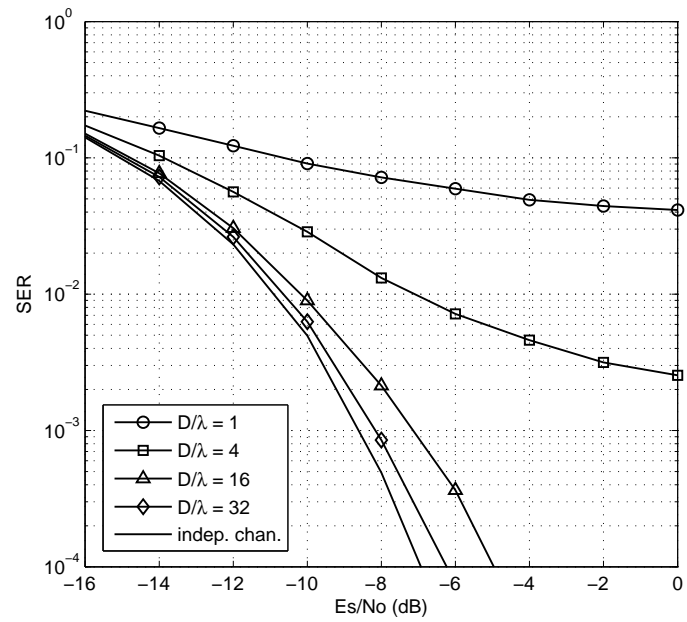}\\
  \caption{SER versus SNR with different array length.}\label{ser_snr_corr_chan}
\end{figure}

Fig.~\ref{ser_snr_corr_chan} shows SER versus SNR when channels are spatially correlated. From the figure, SER improves by increasing the array length since the tap gains corresponding to different antennas are less correlated for larger array length, which coincides with our analysis in Section IV.

\begin{figure}
  \centering
  \includegraphics[width=4in]{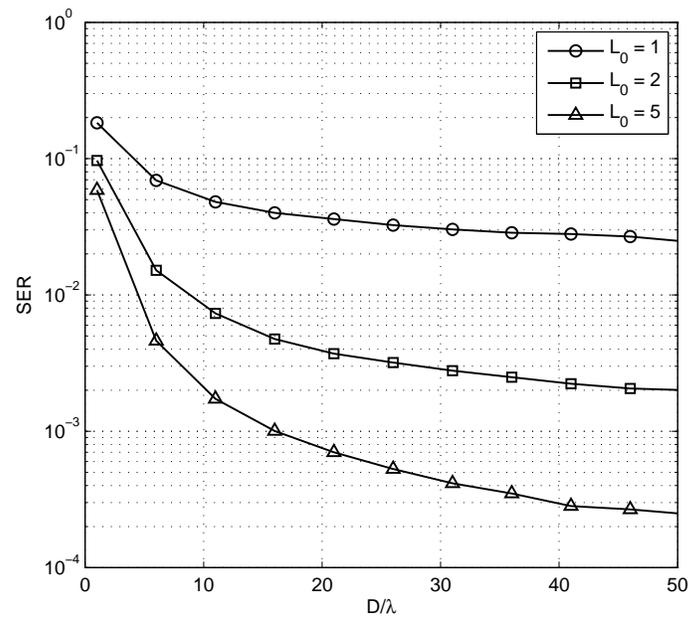}\\
  \caption{SER versus array length with $E_{\mathrm{s}}/N_0=-8$ dB.}\label{ser_dist_corr_chan}
\end{figure}

Fig.~\ref{ser_dist_corr_chan} shows SER versus array length. From the figure, SER can be improved by increasing the array length. When $D=50\lambda$, the SER can be hardly improved by further increasing the array length since the tap gains corresponding to different antennas are sufficiently independent. This corresponds to a distance of $\lambda/2$ between adjacent antennas, which is exactly the Nyquist sample distance in space domain \cite{VTrees}.\par

\begin{figure}
  \centering
  \includegraphics[width=4in]{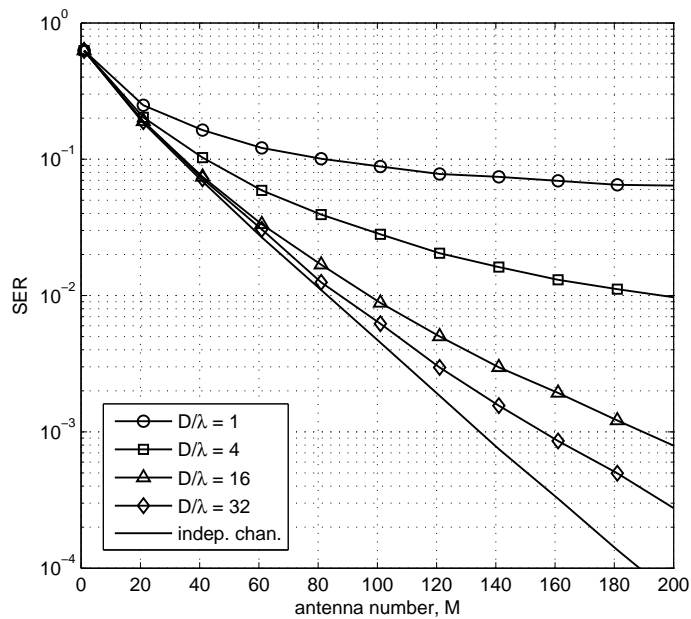}\\
  \caption{SER versus antenna number with $E_{\mathrm{s}}/N_0=-10$ dB.}\label{ser_antnum_corr_chan}
\end{figure}

Fig.~\ref{ser_antnum_corr_chan} shows SER versus antenna numbers with $E_s/N_0=-10$ dB. From the figure, SER reduces as the increasing of the antenna numbers. However, severe performance degradation can be observed due to spatial correlation.

\section{Conclusions}
We have investigated SC modulation for the LSA system in this paper. The LSA-SC receiver has been derived based on the waveform recovery theory, which can provide a straightforward way to understand the behavior of the SC receiver in the large-scale regime. When the antenna number is infinite and the channels at different antennas are independent, the desired waveform can be recovered except for the additive noise and the information-bearing symbols can be thus obtained by sampling the recovered waveform directly. We have also analyzed the non-ideal environments where the the antenna number is finite or the channels at different antennas are correlated. Our work shows that the LSA-SC is a good combination of LSA and SC, and it will be a competitive candidate for the uplink in future wireless systems.
\renewcommand{\theequation}{A.\arabic{equation}}
\setcounter{equation}{0}
\section{Derivation of $\mathrm{E}\left\{|p(t)|^2\right\}$}
Assuming the transmit symbols are independent, $P_{\mathrm{ISI}}$ can be given by
\begin{align}\label{A1}
P_{\mathrm{ISI}}=\sum_{n\neq 0}\mathrm{E}\left\{|f[n]|^2\right\}.
\end{align}

From (\ref{5-4}), $\mathrm{E}\left\{|f[n]|^2\right\}$ can be expressed by
\begin{align}\label{C23}
&\mathrm{E}\left\{|f[n]|^2\right\}\nonumber\\
=&\frac{1}{M^2}\sum_{p_1}\sum_{l_1}\sum_{p_2}\sum_{l_2}g(nT-\tau_{l_1}+\tau_{p_1})g(nT-\tau_{l_2}+\tau_{p_2})\cdot\nonumber\\
&\sum_{m_1=1}^{M}\sum_{m_2=1}^M\mathrm{E}\{\alpha_{m_1}[l_1]\alpha_{m_1}^*[p_1]\alpha_{m_2}^*[l_2]\alpha_{m_2}[p_2]\}.
\end{align}
Using the moment and cumulant relation \cite{CLNikias}, we have
\begin{align}\label{C24-1}
&\mathrm{E}\{\alpha_{m_1}[l_1]\alpha_{m_1}^*[p_1]\alpha_{m_2}^*[l_2]\alpha_{m_2}[p_2]\}=\nonumber\\
&~~~~\mathrm{Cum}\{\alpha_{m_1}[l_1],\alpha_{m_1}^*[p_1],\alpha_{m_2}^*[l_2],\alpha_{m_2}[p_2]\}+\mathrm{E}\{\alpha_{m_1}[l_1]\alpha_{m_1}^*[p_1]\}\mathrm{E}\{\alpha_{m_2}^*[l_2]\alpha_{m_2}[p_2]\}+\nonumber\\
&~~~~\mathrm{E}\{\alpha_{m_1}[l_1]\alpha_{m_2}^*[l_2]\}\mathrm{E}\{\alpha_{m_1}^*[p_1]\alpha_{m_2}[p_2]\}+\mathrm{E}\{\alpha_{m_1}[l_1]\alpha_{m_2}[p_2]\}\mathrm{E}\{\alpha_{m_1}^*[p_1]\alpha_{m_2}^*[l_2]\}.
\end{align}
Since $\alpha_{m_1}[l_1],\alpha_{m_1}^*[p_1],\alpha_{m_2}^*[l_2],\alpha_{m_2}[p_2]$ are Gaussian distributed random variables, we have
\begin{align}
\mathrm{Cum}\{\alpha_{m_1}[l_1],\alpha_{m_1}^*[p_1],\alpha_{m_2}^*[l_2],\alpha_{m_2}[p_2]\}=0,
\end{align}
and thus
\begin{align}\label{C24-1}
&\mathrm{E}\{\alpha_{m_1}[l_1]\alpha_{m_1}^*[p_1]\alpha_{m_2}^*[l_2]\alpha_{m_2}[p_2]\}=\nonumber\\
&~~~~\mathrm{E}\{\alpha_{m_1}[l_1]\alpha_{m_1}^*[p_1]\}\mathrm{E}\{\alpha_{m_2}^*[l_2]\alpha_{m_2}[p_2]\}+\nonumber\\
&~~~~\mathrm{E}\{\alpha_{m_1}[l_1]\alpha_{m_2}^*[l_2]\}\mathrm{E}\{\alpha_{m_1}^*[p_1]\alpha_{m_2}[p_2]\}+\nonumber\\
&~~~~\mathrm{E}\{\alpha_{m_1}[l_1]\alpha_{m_2}[p_2]\}\mathrm{E}\{\alpha_{m_1}^*[p_1]\alpha_{m_2}^*[l_2]\}.
\end{align}
From (\ref{5-2}), we have
\begin{align}
&\mathrm{E}\{\alpha_{m_1}[l_1]\alpha_{m_1}^*[p_1]\}\mathrm{E}\{\alpha_{m_2}^*[l_2]\alpha_{m_2}[p_2]\}=\sigma_{l_1}^2\sigma_{l_2}^2\delta[l_1-p_1]\delta[l_2-p_2],\label{AP1}\\
&\mathrm{E}\{\alpha_{m_1}[l_1]\alpha_{m_2}^*[l_2]\}\mathrm{E}\{\alpha_{m_1}^*[p_1]\alpha_{m_2}[p_2]\}=\sigma_{l_1}^2\sigma_{p_1}^2\rho^2\left(\frac{m_1-m_2}{M-1}\cdot\frac{D}{\lambda}\right)\delta[l_1-l_2]\delta[p_1-p_2].\label{AP2}
\end{align}
Meanwhile, note that
\begin{align}
\mathrm{E}\{\alpha_{m_1}[l_1]\alpha_{m_2}[p_2]\}=\mathrm{E}\{\alpha_{m_1}^*[p_1]\alpha_{m_2}^*[l_2]\}=0.\label{AP3}
\end{align}
Substituting (\ref{AP1}) to (\ref{AP3}) into (\ref{C24-1}),
\begin{align}\label{C25}
&\mathrm{E}\{\alpha_{m_1}[l_1]\alpha_{m_1}^*[p_1]\alpha_{m_2}^*[l_2]\alpha_{m_2}[p_2]\}\nonumber\\
=&\sigma_{l_1}^2\sigma_{l_2}^2\delta[l_1-p_1]\delta[l_2-p_2]+\sigma_{l_1}^2\sigma_{p_1}^2\rho^2\left(\frac{m_1-m_2}{M-1}\cdot\frac{D}{\lambda}\right)\delta[l_1-l_2]\delta[p_1-p_2].
\end{align}
and thus
\begin{align}\label{C26}
\mathrm{E}\left\{|f[n]|^2\right\}=g^2(nT)+\frac{\mathrm{tr}\{\mathbf{R}^2\}}{M^2}\sum_{p}\sum_{l}\sigma_p^2\sigma_l^2g^2(nT-\tau_{l}+\tau_{p}),
\end{align}
where we used the identity
\begin{align}
\sum_{m_1=1}^{M}\sum_{m_2=1}^M\rho^2\left(\frac{m_1-m_2}{M-1}\cdot\frac{D}{\lambda}\right)=\mathrm{tr}\{\mathbf{R}^2\},
\end{align}
for equation (\ref{C26}).\par
Finally, substituting (\ref{C26}) into (\ref{A1}), $P_{\mathrm{ISI}}$ is obtained as
\begin{align}\label{C29}
P_{\mathrm{ISI}}=\frac{\mathrm{tr}\{\mathbf{R}^2\}}{M^2}\sum_{n\neq 0}\sum_{p}\sum_l\sigma_p^2\sigma_l^2g^2(nT-\tau_l+\tau_p),
\end{align}
since $g(nT)=0$ for $n\neq 0$, which is exactly (\ref{6-2}).

\bibliographystyle{IEEEtran}
\bibliography{IEEEabrv,lsmimobib}

\end{document}